\documentstyle[12pt]{article}
\begin{document}
\begin{titlepage}

\flushright{To Appear: {\em Phys. Rev.} {\bf D15}}

\vspace{1in}

\begin{center}
\Large
{\bf Quantum cosmology of generalized two-dimensional dilaton-gravity
models}

\vspace{1in}

\normalsize

\large{James E. Lidsey}

\normalsize
\vspace{.7in}

{\em Astronomy Unit, School of Mathematical
Sciences,  \\
Queen Mary \& Westfield, Mile End Road, LONDON, E1 4NS, U.K.}

\end{center}

\vspace{1in}

\baselineskip=24pt
\begin{abstract}
\noindent

The quantum cosmology of two-dimensional dilaton-gravity
models is investigated. A class of models is  mapped onto the
constrained oscillator-ghost-oscillator model.
A number of exact and approximate solutions to the corresponding
Wheeler-DeWitt equation are presented.
A wider class of minisuperspace models that can be solved in this fashion
is identified.
Supersymmetric extensions to the induced gravity
theory and the bosonic string theory are
then considered and closed-form solutions
to the associated quantum constraints are derived.
The possibility of applying the third-quantization procedure to two-dimensional
dilaton-gravity is briefly discussed.

\end{abstract}

\vspace{.7in}


\end{titlepage}

%\double

\section{Introduction}

\setcounter{equation}{0}

\def\theequation{\thesection.\arabic{equation}}

It is widely thought that
quantum gravitational effects become important on
scales of the order of the Planck length. It follows, therefore,
that insight into the nature of quantum gravity might be
gained  by considering the very early Universe and the
end-point of black hole evaporation.
However, there remain many unresolved technical and conceptual
difficulties with $(3+1)$-dimensional quantum gravity and
to make progress  one must consider simplified toy models.

Recently
there has been considerable interest in two-dimensional theories of gravity.
The Einstein action is  a topological invariant in two dimensions and must
therefore be modified if the theory is to be
non-trivial. The simplest extension is to include a
non-minimally coupled, self-interacting scalar `dilaton' field.
These models are closey related to string theory in non-critical dimensions
\cite{string}
and may provide a solvable framework in which some of the
questions raised in quantum gravity  can be studied. Indeed,
black hole evaporation has been extensively
investigated  following the
introduction of the  Callan, Giddings, Harvey and Strominger (CGHS) model
\cite{CGHS}.

Quantization of two-dimensional models has been performed by employing
a number of techniques such as the BRST and Dirac operator
methods \cite{Banks,dd}. However,
it is possible that lower-dimensional gravity may also shed  light
on some of the issues raised  in quantum cosmology, such as the problem
of extracting physical predictions from the wavefunction
of the Universe.
In this paper we investigate the quantum cosmology
of a generalized class of two-dimensional models.

A number of approaches to quantum cosmology may be taken.
In the path integral formalism, for example,  the wavefunction of the
Universe is expressed as a path integral over
a certain class of metrics, matter distributions and manifolds
\cite{HH,halliwellreview}. This formalism  has recently been investigated
within the context of two-dimensional quantum gravity  by Ishikawa \cite{I}.

Alternatively,  one may follow the Dirac
quantization procedure \cite{dirac}. The wavefunction
of the Universe is annihilated by the Hamiltonian
operator and, in principle, its functional form can be  determined
by solving the zero-energy Schr\"odinger equation.
This equation decomposes into the Wheeler-DeWitt and momentum constraint
equations that describe, respectively,   the invariance of the theory under
reparametrizations of time and spatial diffeomorphisms \cite{WDWmom}.

In general, the Wheeler-DeWitt
equation is a functional differential equation and
is very difficult to  solve. Solutions
can be found in four-dimensional theories
by invoking the `minisuperspace' approximation and  freezing out
the inhomogeneous modes. It is not yet established that
such an approach can lead to meaningful results in higher dimensions,
although arguments have been developed
to suggest that it may be relevant \cite{rel}.
On the other hand, the existence of a Killing vector in a
wide class of spatially closed, two-dimensional cosmologies
implies that all classical
solutions to the field equations are spatially homogeneous \cite{Banks}.
It can
therefore be argued that the minisuperspace approach is exact for
these models \cite{aa,ff}.

Adi and Solomon have adopted a geometrical approach and found a new solution
of the Wheeler-DeWitt equation \cite{ee}.
Navarro-Salas {\em et al.} \cite{aa}
have quantized the induced gravity theory \cite{JT}
via the covariant phase-space and reduced ADM phace-space methods.
They also derived and solved  the Wheeler-DeWitt equation for this model.
Henneaux, on the other hand,  performed the quantization of this theory
in the functional Schr\"odinger representation by first solving
the supermomentum constraint at the classical level \cite{hen}.
This technique was subsequently generalized  to other models \cite{gg}.

In this paper we follow the
approach normally employed in four-dimensional quantum cosmology.
We consider spatially closed cosmologies within the generalized class of
two-dimensional dilaton-gravity models, where the dilaton field
is assumed to be  constant on the surfaces of homogeneity.
The models are defined  by the functional form of the dilaton potential and
specific models have been considered previously in this fashion
by a number of authors \cite{aa,OT,NT,MR1993}.
We find that the Wheeler-DeWitt  equation is exactly solvable
for a wide class of potentials.

The paper is organized as follows.
The classical dynamics of these models is considered in Section 2
and a  subset is mapped onto the
constrained oscillator-ghost-oscillator model.
In Section 3,  these models are quantized and a number
of exact solutions to the Wheeler-DeWitt equation are presented. An
interpretation of these solutions  is discussed within the context of a
specific model. Two classes of approximate solutions
are presented in Section 4. The first is a power series solution to
the Wheeler-DeWitt equation derived by employing
a modification of the Picard iteration scheme \cite{Pic}.
The second is the class of WKB solutions derived by means of a
Legendre transformation \cite{leg}. In Section 5,
a wider class of exactly solvable two-dimensional minisuperspaces
is identified. If appropriate conditions are
satisfied, these models can also be mapped
onto the constrained oscillator-ghost-oscillator model. The superpotential
of the wavefunction must be a separable function of the minisuperspace
null coordinates. We employ this observation to transform the Wheeler-DeWitt
equation derived from a renormalizable dilaton-gravity model
into a solvable form \cite{MR1993}. In Section 6,
a supersymmetric extension is considered. It is found that
the superspace Hamiltonians derived from the induced gravity theory
and the string effective action may each be viewed as
 the bosonic component of a supersymmetric
Hamiltonian. This symmetry is preserved at the quantum level and
the associated quantum constraints are solved exactly in closed form for
both models.  We conclude in Section 7 with a brief
discussion on the possibility of applying the third quantization
procedure to two-dimensional cosmologies.

Units are chosen such that $\hbar =c=1$ unless otherwise stated.

\section{Two-dimensional dilaton-gravity models}

\setcounter{equation}{0}

\def\theequation{\thesection.\arabic{equation}}

\subsection{The generalized action}

The most general
action for two-dimensional dilaton-gravity
that is invariant under local reparametrizations
and does not contain third or higher-order
derivatives is \cite{Banks,RT}
\begin{equation}
\label{action2d}
S=-\frac{1}{2\pi} \int d^2x\sqrt{-g} \left[ c_2(\Phi ) R+c_1(\Phi)
g^{\mu\nu}\partial_{\mu}\Phi \partial_{\nu} \Phi +U(\Phi )\right] ,
\end{equation}
where $g_{\mu\nu}$ is the metric
on the two-dimensional space-time manifold,
$g$ is its determinant, $c_1(\Phi)$ and $c_2 (\Phi)$ are
functions
of the dilaton scalar field $\Phi$, $R$  is the curvature scalar and
$U(\Phi)$ is the dilaton potential.

Specific forms for the functions $\{c_1,c_2,U\}$ correspond
to different two-dimensional models. For example, the induced gravity
action is a special case
of Eq. (\ref{action2d}) with $c_1=1$, $c_2=2\Phi$ and $U=4\lambda^2$
\cite{P}.
 The constant curvature condition may be derived from this
action and  provides a suitable analogue to
the Einstein field equations in two dimensions  \cite{JT}.
Also of interest is the spherically-symmetric
four-dimensional Einstein-Hilbert action. This is equivalent to
Eq. (\ref{action2d}) if
  $c_2=e^{-2\Phi}$, $c_1=2c_2$ and $U=2(1-\Lambda
e^{-2\Phi})$,
where $\Lambda$ is the four-dimensional cosmological constant.
In this example the
 dilaton field is related to the radius of the two-sphere.

Actions  of the form (\ref{action2d}) also arise in string theory
\cite{string}.
To leading order in the inverse string tension $\alpha'$, the tree-level
effective action  for the closed bosonic string in two dimensions
is given by
\begin{equation}
\label{stringaction}
S=-\frac{1}{2\pi} \int d^2x \sqrt{-g} e^{-2\Phi}
\left[ R +4 \left( \nabla \Phi \right)^2 + D(\Phi) \right] ,
\end{equation}
where $D(\Phi)=c=16/\alpha' $ is proportional to the effective
central charge and the tachyon field is assumed to be zero
\cite{perry}.  The
field strength $H_{\mu\nu\lambda}$
of the antisymmetric tensor field vanishes identically
in two dimensions.
This action is closely related to the gravitational
sector of the  CGHS model \cite{CGHS}.

Closed string loop corrections introduce additional terms into the
beta functions and therefore modify the effective action
(\ref{stringaction}) \cite{LF}. If field derivatives are neglected,
only the dilaton potential $D(\Phi)$ is altered and, in general,
the loop corrections have the form
\begin{equation}
\label{loopcorrection}
D(\Phi) = \sum_{n\ge 0} a_n e^{2n\Phi}  ,
\end{equation}
where $a_0=c$ and $n$ represents the number of handles inserted
\cite{Callan}. The values of the other coefficients $a_n$ are
determined by the string theory.

McGuigan, Nappi and Yost have considered
two-dimensional string theories containing gauge fields
\cite{MGY}. They showed that
open strings are governed by the Born-Infeld action for non-linear
electrodynamics \cite{reference}.
Orientable open strings may couple to $SU(N)$ and non-orientable strings
to $SO(N)$ or $Sp(N)$, where $N$ must be even for non-orientable
strings \cite{28}.
When hole and crosscap corrections \cite{anotherreference} are included,
it is found that the modified equations
of motion may be derived from an effective action of the form
(\ref{stringaction}), where
\begin{equation}
D(\Phi) = c- \kappa (N+2\eta )e^{\Phi} ,
\end{equation}
and $\eta =-1,0,+1$ when the gauge group is
$SO(N)$, $SU(N)$, or $Sp(N)$, respectively. $\kappa$ is a
positive-definite open string coupling constant.

Mignemi  has recently investigated
 action (\ref{stringaction}) with
$D=\Lambda e^{-2h\Phi}$ for some arbitrary
constant  $h$, and has found
black hole solutions \cite{rapid}.
This action reduces to the string effective action
in the limit $h\rightarrow 0$ and is also conformally equivalent to
a two-dimensional higher-order,  pure gravity theory with
a Lagrangian given by $L=R^{h/(h-1)}$, where $h\ne 1$ \cite{higher}.

 The general action (\ref{action2d})
 may be simplified after suitable  redefinitions  of the
dilaton and graviton fields \cite{RT}. If  $c_1$ and $c_2$
are positive-definite functions,  we may define a new
scalar field
\begin{equation}
\Sigma \equiv \sqrt{2} \int d\Phi \sqrt{c_1 (\Phi)}
\end{equation}
and perform the conformal transformation
\begin{equation}
\label{conformal}
\tilde{g}_{\mu\nu}\equiv \Omega^2 g_{\mu\nu}, \qquad \Omega^2 \equiv
e^{-2\rho} ,
\end{equation}
where
\begin{equation}
\rho = \frac{c_2}{q^2}-\frac{1}{4} \int^{\Sigma}  d\Sigma ' \left(
\frac{dc_2}{d\Sigma '} \right)^{-1}
\end{equation}
and  $q$ is a constant. The action (\ref{action2d}) transforms to
\begin{equation}
\label{conformalaction}
S=-\frac{1}{2\pi} \int d^2x \sqrt{-\tilde{g}} \left[ \frac{1}{2} q\psi
\tilde{R} +\frac{1}{2} \tilde{g}^{\mu\nu} \partial_{\mu} \psi \partial_{\nu}
\psi + V(\psi) \right]  ,
\end{equation}
where
\begin{equation}
q \psi \equiv 2c_2 [ \Phi (\Sigma )], \qquad V (\psi) =e^{2\rho}U(\Phi)
{}.
\end{equation}
Thus, the models are defined uniquely by the functional form
of the dilaton potential. Unless otherwise stated,
 we shall view theory (\ref{conformalaction}) as our
starting action and we
therefore drop the tildes for notational simplicity.

\subsection{Two-dimensional cosmologies}

In the canonical framework the topology of  space-time
 is  $\Sigma
\times \Re$, where the real line $\Re$ corresponds to the time
dimension and the spatial section $\Sigma$ is either a real line or a
circle $S^1$. The former case applies to two-dimensional
 black hole solutions,
whereas the compact spatial topology is relevant for cosmological
models. In this case the world-interval  has the form
\begin{equation}
\label{line}
ds^2=-N^2(t)dt^2+a^2(t)dx^2  ,
\end{equation}
where $a(t)$ is the radius of the spatial hypersurfaces and $N(t)$
is the lapse function. The spatial
sections represent surfaces of constant $\psi$. When the line element
is given by Eq. (\ref{line}),  the Ricci curvature
satisfies $\sqrt{-g}R=-2\partial_t(aK)$, where the extrinsic
curvature scalar is given by $K=-\dot{a}/(aN)$ and a dot denotes
differentiation with respect to $t$.
It follows, therefore, that
the action (\ref{conformalaction}) takes the form
\begin{equation}
\label{S}
S=\int dt \left[ \frac{1}{N} \left( q \dot{\psi}\dot{a}
+\frac{1}{2} a\dot{\psi}^2 \right) -NaV \right]
\end{equation}
after integration over the spatial sections.

We now proceed to express  the kinetic
terms of Eq. (\ref{S})  in canonical and diagonal forms
by introducing new variables. This will
allow the classical field equations to be solved for appropriate
choices of lapse function and leads to simple forms for the Hamiltonian
constraint.
We begin by  defining  the new coordinate pair
\begin{equation}
\label{uvdefinition}
u\equiv \sqrt{2}q a e^{\psi/2q} , \qquad v\equiv \sqrt{2}q  e^{-\psi/2q} .
\end{equation}
The range of these variables is determined
by physical considerations.
The physically interesting region of parameter space is
$0 < \{a,q\psi \}< +\infty$ and this corresponds to the range
 $0<u<+\infty$
and $0<v<\sqrt{2}q$. In terms of these variables
the action (\ref{S}) has the form
\begin{equation}
\label{S1}
S=-\int dt \left[ \frac{1}{N} \dot{u}\dot{v} +NaV \right]
\end{equation}
and the Hamiltonian constraint, derived by functionally differentiating
the action with respect to the non-dynamical  lapse function, is given
by
\begin{equation}
\label{Hequation}
\frac{1}{N^2} \dot{u}\dot{v} -\frac{uv}{2q^2} V[\psi (v)] =0  .
\end{equation}

The second term
on the left-hand side is the `superpotential'.
Its direct dependence on the
dynamical degrees of freedom may
be eliminated by introducing the rescaled
variables
\begin{equation}
\label{alphabeta}
\alpha \equiv \frac{u^2}{4q^2}
\end{equation}
and
\begin{equation}
\label{betaalpha}
\beta \equiv \int^v dv'v' V[\psi (v') ]
=-q \int^{\psi}  d\psi ' e^{-\psi ' /q} V(\psi ') ,
\end{equation}
where $\beta$ represents a rescaled dilaton field.
It follows that $0<\alpha <+\infty$, but the range of values
spanned by $\beta$ is model-dependent.
 The Jacobian of this transformation
vanishes if $V(\psi )$  vanishes and we  therefore restrict
our discussion to potentials that are either positive- or
negative-definite. The action (\ref{S1}) now takes the form
\begin{equation}
\label{Salpha}
S=-\int dt \left[ \frac{1}{N aV}  \dot{\alpha}\dot{\beta} +
 NaV \right]
\end{equation}
and the corresponding Hamiltonian
constraint (\ref{Hequation}) becomes
\begin{equation}
\label{constraint}
H=aV(p_{\alpha} p_{\beta} -1) =0 ,
\end{equation}
where $p_{\alpha}=-\dot{\beta}/(NaV)$
and $p_{\beta} =-\dot{\alpha}/(NaV)$ are
the momenta conjugate to $\alpha$ and $\beta$,  respectively.

In the gauge $N^{-1} =aV$,
the field equations take  the simple form
 $\ddot{\alpha}=\ddot{\beta}=0$
and have the general solution
\begin{eqnarray}
\alpha =\alpha_0+b (t-t_0) \nonumber \\
\beta =\beta_0+ b^{-1} (t-t_0) ,
\end{eqnarray}
where $\{ \alpha_0 , \beta_0 ,t_0 ,b \}$ are constants.
We conclude, therefore, that
the classical dynamics of these two-dimensional
Universes is equivalent to that of a non-interacting point particle
propagating on two-dimensional Minkowski space.
The variables $\alpha$ and $\beta$ may be viewed
as null coordinates and the regime of  Minkowski space accessible
to the `particle' depends on the dilaton potential.

We now impose the additional restriction
that $\beta$ remains positive- or negative-definite for
all physically relevant values of
the scale factor and dilaton field. In this case
we may introduce a third pair of variables defined by\footnote{This
restriction on $\beta$ ensures
that the corresponding Wheeler-DeWitt equation will not be
of the elliptic type
when expressed in terms of $w$ and $z$.}
\begin{equation}
\label{wzdef}
w\equiv \sqrt{\gamma\beta} -\sqrt{\alpha}, \qquad z\equiv
\sqrt{\gamma\beta} +\sqrt{\alpha}  ,
\end{equation}
where $\gamma =\beta/|\beta |$. For real values of $\alpha$ and $\beta$,
$z\ge |w|$. In the gauge
\begin{equation}
N=\frac{1}{aV}  (\alpha \gamma \beta)^{1/2} ,
\end{equation}
action (\ref{Salpha}) transforms to
\begin{equation}
\label{hamact}
S= -\int dt \left[  \gamma \left(  \dot{z}^2 -\dot{w}^2 \right)
+\frac{1}{4} \left( z^2 -w^2 \right) \right] .
\end{equation}
This is the action for the  constrained oscillator-ghost-oscillator
system when $\beta <0$ and the
model corresponds to a constrained hyperbolic system if $\beta >0$.
In this latter case, the field
equations have the general solution
\begin{eqnarray}
w= Ae^{t/2} +Be^{-t/2} \nonumber \\
z=Ce^{t/2} +De^{-t/2} ,
\end{eqnarray}
where the constants of proportionality
satisfy $AB=CD$ and are chosen to ensure $z\ge |w|$ for all $t$.
The trajectory of this solution is a central
conic section in this  sector of the $(w,z)$ plane. If $\beta<0$, however,
the general solution is
\begin{equation}
w=\epsilon E \cos \left( t/2 +\theta_1 \right) ,\qquad
z=E \cos \left( t/2 +\theta_2 \right) ,
\end{equation}
where $\{ E,\theta_1 ,\theta_2 \}$ are
arbitrary, real constants
and $|\epsilon | =1$.
These solutions lie on the family of ellipses \cite{OT}
\begin{equation}
\label{family}
w^2+z^2 -2\epsilon wz \cos \theta =E^2 \sin^2 \theta  ,
\end{equation}
where the eccentricity is determined by $\theta \equiv \theta_1-\theta_2$ and
the major axis lies along the line $w=\epsilon z$.

This correspondence between
the constrained oscillator-ghost-oscillator
and a specific two-dimensional dilaton-gravity model
was recently observed
by \"Onder and Tucker \cite{OT} in the synchronous gauge $N=1$. Their model
corresponds to the choice $c_1=c$, $c_2=\frac{1}{2} \Phi$ and
$U=\Lambda +\lambda  e^{c\Phi}$ in action (\ref{action2d}),
where $\{c, \lambda , \Lambda \}$ are  constants. Within the context
of this model, they employed such
a correspondence to investigate the connection between
the  classical and quantum cosmologies. They  identified
appropriate linear superpositions of quantum states that
highlighted the classical orbits (\ref{family})
and were therefore able to conclude that
a definite correlation between classical and quantum
solutions exists in this model.  This is interesting because the question
of how a classical space-time emerges from a quantum theory
of the Universe is currently unresolved.
The above analysis generalizes the results of Ref. \cite{OT}
and shows that the correspondence  between the
oscillator model and two-dimensional cosmologies
arises in a wide class of dilaton-gravity models.
This suggests that two-dimensional theories may provide valuable insight
into the problems associated with quantum cosmology in higher dimensions.
In view of this we proceed in the next Section to
investigate the quantum cosmological behaviour of
these models.

\section{Exact quantum wavefunctions}

\setcounter{equation}{0}

\def\theequation{\thesection.\arabic{equation}}

\subsection{The Wheeler-DeWitt equation}

The cosmological models defined by Eq. (\ref{Salpha}) are  quantized
with the algebra $[\alpha ,p_{\alpha} ]_- =i$ and
$[\beta ,p_{\beta} ]_- =i$.
The Wheeler-DeWitt equation is the operator
form of the Hamiltonian constraint (\ref{constraint})  and is realized by
identifying $p_{\alpha}$ with $-i\partial /\partial \alpha$ and $p_{\beta}$
with $ -i \partial /\partial \beta $. The physical states,
$\Psi$, of the Universe are annihilated by this Hamiltonian operator.
We shall not
consider the ambiguities that arise in operator ordering, so
the Wheeler-DeWitt equation has the form
\begin{equation}
\label{WDWab}
\left[ \frac{\partial^2}{\partial \alpha \partial \beta} +1 \right] \Psi =0
{}.
\end{equation}
This equation admits a number of exact solutions \cite{page}.
One family is given by
\begin{equation}
\label{absol}
\Psi_b =e^{-ib\alpha -i \beta /b} ,
\end{equation}
where $b$ is an arbitrary,  complex constant. $|\Psi_b |$
is bounded everywhere when  ${\rm Im}b=0$ and is
bounded for ${\rm Im} b <0$ if  $\beta <0$.

A natural generalization of this solution is
 to include a variable amplitude $\Delta (\alpha ,\beta )$.
Substitution of the ansatz $\Psi =\Delta \Psi_b$ into Eq. (\ref{WDWab})
implies that $\Delta$ satisfies
\begin{equation}
\frac{\partial^2 \Delta}{\partial \alpha \partial \beta} -\frac{i}{b}
\frac{\partial \Delta}{\partial \alpha} -ib \frac{\partial \Delta}{\partial
\beta} =0
\end{equation}
and one non-trivial solution to this equation is
\begin{equation}
\Delta={b} \alpha -\frac{\beta}{b} .
\end{equation}

In terms of the coordinate pair (\ref{wzdef})
the Wheeler-DeWitt equation (\ref{WDWab}) transforms
into \cite{OT,page,lidsey,lots,HP1990,CG}
\begin{equation}
\label{WDWwz}
\left[ \frac{\partial^2}{\partial w^2} -\frac{\partial^2}{\partial z^2} +
\gamma \left( w^2-z^2 \right) \right] \Psi =0 .
\end{equation}
This has separable solutions of the form $\Psi = \sum_n c_n \Psi_n$,
where
\begin{equation}
\label{wzsol}
\Psi_n =H_n \left[ \sqrt{c} w \right] H_n
\left[ \sqrt{c} z\right]
e^{-c \left[ w^2 +z^2 \right]/2} ,
\end{equation}
$c_n$ are arbitrary
complex coefficients,
$H_n$ is the Hermite polynomial of order $n$ and $c=i$
$(c=1)$ for $ \gamma =+1$ $( \gamma =-1)$.

If $\beta$ does not change sign, a
 third class of  solution is generated
by defining
the variables
\begin{eqnarray}
\label{srdef}
s=\frac{1}{6} \ln \left( 4\alpha \gamma \beta  \right) \nonumber \\
r=\frac{1}{6} \ln \left( \frac{\alpha}{\gamma\beta} \right)
\end{eqnarray}
and  the Wheeler-DeWitt equation becomes
\begin{equation}
\label{WDWsr}
\left[ \frac{\partial^2}{\partial s^2} -\frac{\partial^2}{\partial r^2} +9
\gamma e^{6s} \right] \Psi =0 .
\end{equation}
The wavefunction is an eigenstate  of the momentum
operator $\partial /\partial r$ and has the separable form
\begin{equation}
\label{solutionsr}
\Psi_p = e^{ipr} Z_{\pm ip/3} \left( \sqrt{ \gamma} e^{3s}
\right) ,
\end{equation}
where $p$ is a separation constant and $Z_{\pm ip/3}$ represents  a linear
combination of ordinary Bessel functions of order $\pm ip/3$.

It should be noted that technical questions
 arise when quantizing  with variables
that are restricted to a finite range, as is the case in the derivation
of Eqs. (\ref{WDWab}) and (\ref{WDWwz}). However, these issues are beyond
the scope of the present work
\cite{tech}. On the other hand,  the variables (\ref{srdef})
 are unrestricted and Eq. (\ref{WDWsr}) can be derived  from
the corresponding classical action with an appropriate choice of factor
ordering. Eqs. (\ref{WDWab}) and (\ref{WDWwz}) may then be derived
directly from this equation by a change of variables.

An additional class of exact solutions may be generated by
defining new variables
\begin{equation}
\label{munudef}
\mu \equiv \frac{\alpha}{2}+\sqrt{2\gamma
\beta}, \qquad \nu \equiv \frac{\alpha}{2} -\sqrt{
2\gamma \beta} .
\end{equation}
The Wheeler-DeWitt  equation transforms into
\begin{equation}
\left[ \frac{\partial^2}{\partial \mu^2} -\frac{\partial^2}{\partial \nu^2} +
\gamma (\mu -\nu ) \right] \Psi =0
\end{equation}
and has the separable solution
\begin{equation}
\label{airys}
\Psi = \left[ c_1 {\rm Ai} (m- \gamma \mu ) +
c_2 {\rm Bi} ( m-\gamma \mu ) \right]
\left[ c_3 {\rm Ai} (m- \gamma \nu ) +c_4 {\rm Bi} (m-
 \gamma  \nu )\right]
\end{equation}
in terms of Airy functions, where $\{m, c_j\}$ are arbitrary
constants \cite{Airy}.

\subsection{Exponential dilaton interactions}

It is useful to consider a specific model in order to discuss
these  solutions.  We shall investigate an
action of the form (\ref{stringaction}), where the dilaton
potential  is given by $D=\Lambda e^{-2h \Phi}$
for some constants $\{ h, \Lambda \}$ \cite{rapid}.
This action is conformally equivalent to Eq. (\ref{conformalaction}), where
\begin{equation}
q \psi = 2e^{-2\Phi}, \qquad
V(\psi )=\frac{\Lambda}{8} \left( \frac{q\psi}{2} \right)^h
e^{\psi /q} .
\end{equation}
The kinetic terms in the action (\ref{S}) may
be diagonalized by introducing the variables
\begin{eqnarray}
\label{XT}
T=\frac{q}{\sqrt{2}} \left( ae^{\psi /2q}+e^{-\psi /2q} \right)
\nonumber \\
X=\frac{q}{\sqrt{2}} \left( ae^{\psi /2q} -e^{-\psi /2q} \right)
\end{eqnarray}
and this implies that the minisuperspace metric is transformed
into the Minkowski metric, where
$T$ is the timelike coordinate and $X$ is
the spacelike coordinate. Since $T\ge |X|$,
only the interior region of the future
light cone of the origin is covered by
this coordinate system.  We deduce, therefore, that
the variables (\ref{uvdefinition}) represent the null coordinates
$u=T+X$ and $v=T-X$ in this region. It should be noted that
only a finite region of the interior of the
future light cone is represented because
$v <v_{\rm max} = \sqrt{2}q$ is bounded from above.

For this model the $(\alpha , \beta )$ coordinates
are given by
\begin{equation}
\label{alphamodel}
\alpha =\frac{1}{2} a^2 e^{\psi /q}
\end{equation}
and
\begin{eqnarray}
\label{betamodel}
\beta =-\frac{\Lambda}{4} \frac{1}{1+h} \left(
\frac{q\psi }{2} \right)^{1+h}, \qquad h\ne -1
\nonumber \\
\beta = -\frac{\Lambda}{4} \ln  (\psi ) ,\qquad h=-1 ,
\end{eqnarray}
respectively, and when
 $h\ne -1$, the sign of $\beta$ is uniquely determined
by the sign of $\Lambda /(1+h)$.

The parameters $\{ \alpha ,\beta \}$  may also be viewed as null coordinates
over a region of Minkowski space spanned by the  timelike
coordinate  $\tilde{p}=\alpha +\beta$ and spacelike
coordinate  $\tilde{q}=\alpha -\beta$. Thus,
if $\beta >0$, it follows that
$\tilde{p}\ge |\tilde{q}|$ and the analysis is
again restricted to the interior of the future
light cone of the origin. Since $\alpha ,\beta \in (0,\infty )$,
the whole of the interior is now covered.
On the other hand, if $\beta <0$, $\tilde{q}\ge |\tilde{p}|$, and
the propagation of the wavefunction  is restricted to the Rindler
wedge of Minkowski space.

The simplest interpretation of the wavefunction identifies
an oscillating solution to the Wheeler-DeWitt equation as a
Lorentzian geometry and a cosmological singularity is
associated with an infinite number of oscillations \cite{H1984}. A
non-oscillating solution represents a classically  forbidden
Euclidean geometry. Let us
 consider the case where $\beta <0$ and $h\ne -1$.
Eq.  (\ref{wzsol}) represents the basis for
a discrete spectrum of Euclidean solutions, where
the parameter $n$ determines the excitation level of the wavefunction
\cite{HP1990}.
The ground state is associated with $n=0$ and excited states with $n>0$.
This ground state is identical to solution (\ref{absol})
when $b=-i$. Hence, we may view the solution $\Psi_{b=-i}$ as the
ground state of a continuous spectrum of excited states (\ref{absol})
that are parametrized by
the separation constant $b$ with  ${\rm Re}b=0$ and ${\rm Im}b<0$.

Although these classes of
Euclidean solution appear to correspond to
 classically forbidden behaviour,
Lorentzian wavefunctions may be generated
from appropriate
linear combinations of the excited states.
A more general solution to Eq. (\ref{WDWab})
is given by \cite{page}
\begin{equation}
\label{generalll}
\Psi=\int_C dc M(c) e^{-c\alpha - \gamma \beta  /c} ,
\end{equation}
where $M(c)$ is an arbitrary  function of the parameter
$c\equiv ib$ and $C$
represents some contour of integration in the complex plane. If
$M(c) =\frac{1}{2} c^{(ip-3)/3}$ and the contour of integration
is over the positive half of the real axis, Eq. (\ref{generalll})
may be evaluated exactly in terms of the modified Bessel function:
\begin{equation}
\Psi_p=\frac{1}{2} \int^{\infty}_0 dc c^{(ip-3)/3} e^{-c
\alpha -|\beta |/c} =  e^{ipr} K_{ip/3}
\left( 2\sqrt{\alpha\gamma\beta} \right) .
\end{equation}
We recognize this superposition  as solution (\ref{solutionsr})
with $Z = K (e^{3s})$.
This solution may also be generated from a linear
combination of harmonic oscillator wavefunctions and, in general,
solutions (\ref{absol}), (\ref{wzsol}) and (\ref{solutionsr})
may be expressed as linear combinations of one other for positive
and negative $\beta$ \cite{page,HP1990,CG}.

The modified Bessel
function has the asymptotic form $K_q(x) \propto x^{-q}$ for $|x|\ll 1$
and $q\ne 0$. Thus, the wavefunction has the form $\Psi_p \propto
e^{ip(r \pm s)}$ for small spatial  geometries $(s \rightarrow -\infty )$
and these  represent plane waves in the variables $(r,s)$.
The wavefunction oscillates an
infinite number of times when the spatial volume of the Universe vanishes
and we identify this point
as a cosmological singularity. However, the wavefunction is exponentially
damped for $e^{3s} > |p|/3$ and this region of minisuperspace
is classically  forbidden.
It is interesting to relate this solution to the classical
solution  in terms of the variables (\ref{srdef}).
The gravitational field equations derived from  action (\ref{Salpha})
in the gauge $N=9e^{6s}/(4aV\gamma)$ are given by
\begin{equation}
\ddot{r} =0, \qquad
\ddot{s} =\frac{27}{4\gamma} e^{6s}
\end{equation}
and the  general solution satisfying the Hamiltonian constraint is
\begin{eqnarray}
r=At+B \nonumber \\
\pm t =\frac{1}{3A} \ln \left[ e^{-3s} + \left(  e^{-6s}
-\frac{9}{4A^2} \right)^{1/2} \right] ,
\end{eqnarray}
where $\{A,B \}$ are arbitrary constants. (We
have perfomed a linear translation on  $t$
without loss of generality).
It follows  that the value of $s$ is bounded from above
by the constraint $e^{3s} <2A/3$ and we may therefore
identify the
eigenvalue of the momentum operator $\partial /\partial r$ with the
integration
constant $A$, i.e. $|p| =2A$.

In general, it is difficult to evaluate Eq. (\ref{generalll})
exactly. However, it can be related directly to solution (\ref{airys})
by performing a trivial rescaling $\tilde{c}=- 2c$ and  specifying
\begin{equation}
M (\tilde{c} )=\frac{1}{\tilde{c}^{1/2}}
\exp \left[ -\frac{{\tilde{c}}^3}{12} +m {\tilde{c}} \right] ,
\end{equation}
where $m$ is an arbitrary constant. Substitution of
variables (\ref{munudef}) into Eq. (\ref{generalll}) therefore implies
that
\begin{equation}
\label{integral}
\Psi = \int_C \frac{d{\tilde{c}}}{{\tilde{c}}^{1/2}} \exp \left[ -
\frac{{\tilde{c}}^3}{12}
+(\mu +\nu +2m) \frac{{\tilde{c}}}{2} +\frac{1}{4{\tilde{c}}} (\mu -\nu )^2
\right] .
\end{equation}
An integral of this form has been
evaluated previously by Halliwell and Louko within
the context of the path integral quantization of the four-dimensional
de  Sitter Universe \cite{HL}. They
showed that it may be expressed in terms
of products of Airy functions, where
the specific combination is determined
by the contour of integration. In our example, we wish to construct
wavefunctions from a linear combination of bounded wavefunctions of
the form (\ref{absol}) and therefore require ${\rm Re }{\tilde{c}}<0$.
If the contour of integration is chosen to
lie along the line ${\tilde{c}} =i\eta -\epsilon$, where $\eta$ is real
and $\epsilon >0$, it will pass to the left of the origin in the complex
plane. In this case, Eq. (\ref{integral}) is given by \cite{HL}
\begin{equation}
\Psi = {\rm Ai} (\mu +m ) {\rm Ai} (\nu +m)  .
\end{equation}
Up to a numerical constant, this solution corresponds
to Eq. (\ref{airys}) with $c_1=c_3$ and $c_2=c_4 =0$.
It oscillates if either $\nu+m<0$ or $\mu +m<0$ and is
exponentially damped when both arguments are positive.

Thus far,  we have derived exact solutions to the
Wheeler-DeWitt equation. However, it is
useful to search for approximate solutions as well. Although
such solutions are not exact, they can provide insight
into the nature of the wavefunction. In the following
Section, we shall discuss two classes of approximate
solutions.

\section{Approximate wavefunctions}

\setcounter{equation}{0}

\def\theequation{\thesection.\arabic{equation}}

\subsection{Power series solutions}

Power series solutions to the Wheeler-DeWitt
equation (\ref{WDWab}) may be derived  by expanding the
wavefunction as the infinite sum of functions
\begin{equation}
\label{Psiseries}
\Psi =\sum_{m=0}^{\infty} \Psi_m .
\end{equation}
This ansatz is a consistent solution to Eq. (\ref{WDWab}) if
\begin{equation}
\label{firstterm}
\frac{\partial^2 \Psi_0}{\partial \alpha\partial \beta} =0
\end{equation}
and
\begin{equation}
\label{otherterms}
\frac{\partial^2 \Psi_m}{\partial \alpha \partial \beta} =-
\Psi_{m-1}, \qquad m\ge 1 .
\end{equation}
Eq. (\ref{firstterm}) is  the canonical, one-dimensional wave
equation and has the general solution
\begin{equation}
\Psi_0 =P(\alpha)+Q(\beta) ,
\end{equation}
where $P$ and $Q$ are arbitrary, twice continuously differentiable
functions. A modification of the Picard iteration scheme \cite{Pic,lidsey}
 may now be established  by expressing $\Psi_m$ in terms of quadratures
with respect to the null variables $\{ \alpha ,\beta \}$.
When $m=1$, Eq. (\ref{otherterms}) admits the separable solution
\begin{equation}
\Psi_1 =-  \left[  \beta \int^{\alpha} d\alpha_1
P(\alpha_1) + \alpha \int^{\beta} d\beta_1 Q(\beta_1) \right]
\end{equation}
and this result may then be substituted back into Eq. (\ref{otherterms})
to derive $\Psi_2$ and so on. The general pattern is easy to deduce
after a few iterations and we conclude, therefore,
 that
\begin{eqnarray}
\label{powersolution}
\Psi =\Psi_0 +\sum_{m=1}^{\infty} \frac{(-1)^m}{m!}
\left[ \beta^m \int^{\alpha} d\alpha_m \ldots \int^{\alpha_3}
d {\alpha}_2 \int^{\alpha_2}
d\alpha_1 P(\alpha_1) \right. \nonumber \\
\left. +\alpha^m \int^{\beta}
d\beta_m \ldots \int^{\beta_3} d\beta_2 \int^{\beta_2} d\beta_1
Q(\beta_1)
\right]
\end{eqnarray}
is also a solution to the Wheeler-DeWitt Eq. (\ref{WDWab}).

It should be emphasized that we have not assumed a semi-classical
approximation in deriving
these power series solutions. The advantage of this scheme
is that the wavefunction is given in terms of {\em arbitrary}
functions of $\alpha $ and $\beta$. In many minisuperspace
models these variables are related to the spatial
volume of the Universe, where small values of $\alpha$ or $\beta$
typically correspond to small spatial volumes. Indeed,  this is the case
for two-dimensional dilaton-gravity cosmologies, since $\alpha$
is proportional to the square of the scale factor. Consequently,
for  $P(\alpha) =0$,   we may  view  solution
(\ref{powersolution}) in the region
of the origin as an expansion in powers of a small parameter $\alpha$.

\subsection{Semi-classical wavefunctions}

Within the context of four-dimensional cosmologies, the nature
of space-time is accurately  described by classical physics when
the spatial volume of the
Universe is significantly larger than the Planck scale.
It follows, therefore, that classical behaviour
from the quantum regime should be predicted  by  the quantum theory.
Presently, the problem of how such a transition might occur
is an unresolved one. However, it is reasonable
to suppose that  the nature of semi-classical wavefunctions may
provide some insight.

In the WKB approximation, corresponding to the limit
$\hbar \rightarrow 0$, one treats
some of the degrees of freedom  $\{ c\}$ as classical variables
and the remainder $\{ q\}$ quantum-mechanically. The wavefunction is then
viewed as a linear superposition of waves
of the form $\Psi \approx e^{-iS/\hbar }$,
where $S$ is the classical action satisfying  the Hamilton-Jacobi equation.
This equation is derived by identifying  the conjugate
momenta in Eq. (\ref{constraint}) with $p_{\alpha}
= \partial S/\partial \alpha$ and $p_{\beta}=\partial S
/\partial \beta$. It takes the form of a non-linear, first-order
partial differential equation:
\begin{equation}
\label{HJ}
\frac{\partial S}{\partial \chi}\frac{\partial S}{\partial \beta} = \chi ,
\end{equation}
where a new variable
$\chi \equiv \sqrt{2\alpha}$ has been introduced.
It is well known that
there exists a one-to-one correspondence between
congruences of classical solutions and solutions to
the Hamilton-Jacobi equation in two-dimensional minisuperspaces
\cite{referee}.
In principle, therefore, an arbitrary solution to Eq. (\ref{HJ}) may
be generated once the classical solutions are known.

However, parametric solutions may be found
more directly by employing a Legendre transformation \cite{leg}. We
define new variables
\begin{equation}
\label{xi eta}
\xi \equiv   \frac{\partial S}{\partial \chi}, \qquad \eta \equiv
\frac{\partial S}{\partial \beta}
\end{equation}
and a new function
\begin{equation}
\label{rho}
\rho (\xi ,\eta ) \equiv \chi \xi +\beta \eta -S(\chi ,\beta ) .
\end{equation}
Partial differentiation with respect to $\xi$ implies that
\begin{equation}
\label{enroute}
\chi =\frac{\partial \rho}{\partial \xi} =  \xi \eta ,
\end{equation}
where the second equality follows from Eq. (\ref{HJ}). Eq. (\ref{enroute})
has the {\em general} solution
\begin{equation}
\label{general}
\rho =\frac{1}{2} \eta \xi^2 +f(\eta) ,
\end{equation}
where $f(\eta)$ is an arbitrary function of $\eta$.

A parametric solution may now be found
 by transforming back into the
old variables. After differentiation of Eq. (\ref{general}) with
respect to $\eta$, we find that
\begin{eqnarray}
\label{parametricsolution}
S=\frac{2 \alpha}{\eta}
-f(\eta) +\eta \frac{df}{d\eta} \nonumber \\
\beta=\frac{\alpha}{\eta^2} +\frac{df}{d\eta} .
\end{eqnarray}

It should be noted that this Legendre transformation is only
self-consistent if the Jacobian
\begin{equation}
\label{jacobian}
J=\frac{\partial^2 S}{\partial \chi^2} \frac{\partial^2 S}{\partial \beta^2}
-\left(
\frac{\partial^2 S}{\partial \chi\partial \beta} \right)^2
\end{equation}
is non-vanishing. Solutions are said to be `developable'
if $J\ne 0$ and `non-developable' if $J=0$. All
developable solutions can be written in the parametric form
of Eq. (\ref{parametricsolution}). In principle,
 we can determine $\eta =\eta
(\alpha ,\beta )$ from the second equation in (\ref{parametricsolution})
once
the functional form of  $f(\eta)$ has been  specified.
Substituting  this  result into the first equation
yields the action in terms of the canonical variables,
or equivalently,  in terms of the original variables
 via Eqs. (\ref{uvdefinition}), (\ref{alphabeta}) and (\ref{betaalpha}).

For example, the Jacobian is non-vanishing if
\begin{equation}
f(\eta ) = \frac{\alpha_i}{\eta} +\beta_i \eta ,
\end{equation}
where $\{\alpha_i,\beta_i\}$ are finite constants, and this
ansatz leads  to the action
\begin{equation}
\label{actionsolution}
S=2\sqrt{\left( \alpha-\alpha_i\right) \left( \beta-\beta_i \right) } .
\end{equation}
When $f=0$, this solution is closely related to the  exact solution of Eq.
 (\ref{WDWsr}) that is given by
$\Psi =H^{(2)}_0 \left( 2\sqrt{\alpha \beta} \right)$,
where $H^{(2)}_0 (x)$
is the Hankel function.  For small arguments this function has
the asymptotic form $ 2i\pi^{-1} \ln (x)$, so  the wavefunction
does not oscillate. On the other hand,
the Hankel function has the form
$H_0^{(2)} (x) \propto  x^{-1/2} e^{\pm  ix}$ at large arguments
and this does have oscillatory
behaviour. In this example, a large argument corresponds to a large
value of the scale factor. Consequently, the argument
of this solution may be identified with the action
(\ref{actionsolution}) and represents  a classically
allowed solution that has tunneled from the Euclidean regime.

Exact solutions  can also be found if $f\propto \eta^{\pm 3}$
and $f \propto \ln \eta$.  Furthermore,
it is interesting to note that solution (\ref{absol}) to the full
Wheeler-DeWitt equation (\ref{WDWab})
is of the WKB form $\Psi =e^{-iS}$, where
$S=b \alpha + b^{-1}\beta $. This
 is an exact, non-developable solution to the
Hamilton-Jacobi equation (\ref{HJ}) and in this sense
 the WKB approximation is exact for this solution.

This concludes our discussion on approximate solutions.
In the next Section we shall investigate whether other
minisuperspace models can be solved in the manner discussed
above.

\section{A class of integrable minisuperspaces}

\setcounter{equation}{0}

\def\theequation{\thesection.\arabic{equation}}

It is interesting to investigate whether a wider class
of models leads to the  Wheeler-DeWitt
equation (\ref{WDWab}). To proceed, we investigate
an equation of the form
\begin{equation}
\label{KG}
\left[
\frac{\partial^2}{\partial x^2} -\frac{\partial^2}{\partial y^2} +4m^2(x ,y)
\right] \Psi =0 ,
\end{equation}
where the superpotential, $m^2(x,y)$, is some function of
the minisuperspace coordinates $(x,y)$.

We introduce new variables $\alpha = \alpha (\sigma)$ and
$\beta = \beta
(\tau )$ that are {\em arbitrary} functions of the minisuperspace
null coordinates $\sigma \equiv x + y$ and $\tau \equiv
x -y$. These new variables satisfy the boundary conditions
$\partial \alpha /\partial x =\partial \alpha /\partial y$ and $\partial
\beta /\partial
x =-\partial \beta  /\partial y$
and these constraints ensure that the derivative terms in Eq.
(\ref{KG}) are transformed into the canonical form:
\begin{equation}
\label{*}
\left[ \frac{\partial \alpha}{\partial x} \frac{\partial \beta}{\partial x}
\frac{\partial^2}{\partial  \alpha  \partial \beta}
 +m^2 \right] \Psi =0 .
\end{equation}

It follows that Eq. (\ref{*}) reduces to Eq. (\ref{WDWab}) if the new
variables $\alpha$ and $\beta$ are themselves solutions to the equation
\begin{equation}
\label{masscondition}
m^2 = \frac{\partial \alpha}{\partial x}\frac{\partial \beta}{\partial x} =
\frac{d\alpha}{d\sigma}
\frac{d\beta}{d\tau} .
\end{equation}
In principle, therefore, Eq. (\ref{KG}) may be solved
if a solution to the constraint equation
(\ref{masscondition}) can be found. Effectively, the problem of
solving the linear, second-order partial differential equation
(\ref{KG}) has been reduced to finding a solution to the non-linear,
{\em first-order} equation (\ref{masscondition}) and in many cases it
is considerably easier to solve this latter equation.
Indeed, it is clear from the second equality in Eq.
(\ref{masscondition}) that when the superpotential has the generic
form
\begin{equation}
\label{null}
m^2 (x , y ) = m_+ (\sigma )m_- (\tau ) ,
\end{equation}
where $m_{\pm}$ are some known analytical functions, Eq.
(\ref{masscondition}) admits the general separable solution
\begin{equation}
\label{uv}
\alpha=\lambda
\int^{\sigma} d\sigma' m_+ (\sigma' ), \qquad \beta = \lambda^{-1}
\int^{\tau} d\tau'
 m_- (\tau') ,
\end{equation}
where $\lambda$ is an arbitrary separation constant. The region
of minisuperspace covered by these coordinates is determined by the
specific form of the  superpotential.

The Jacobian of the transformation leading to Eq. (\ref{*}) vanishes
whenever the null variables
$\alpha =\alpha (\sigma )$ or $\beta =\beta (\tau )$
have turning points and
these will occur at the zero points of the superpotential if
Eq. (\ref{masscondition}) is satisfied.
Thus, the Wheeler-DeWitt equation can be mapped onto
the unit-mass Klein-Gordon equation  if the superpotential
is positive- or negative-definite over
the whole region of minisuperspace covered by $\{ \alpha ,
\beta \}$ and, in addition,
is a  separable function  of
these null coordinates. For example,  if  $\alpha >0$ and $\beta <0$,
the  solutions discussed in Section 3.2 are also solutions
to Eq. (\ref{KG}).
If, on the other hand,  the superpotential does vanish at
some point in minisuperspace,
 equivalent transformations to those discussed above may
be performed on both sides of the zero-point. The two solutions
in the different regions may then be matched at the boundary by
requiring that the wavefunction and its first derivative are continuous
\cite{Uglum}.

There are a number of interesting minisuperspaces for which Eq.
(\ref{masscondition}) can be solved exactly. In many cases the
superpotential of the wavefunction is independent of one of the null
coordinates, i.e. it is a single function of either $\sigma$ or
$\tau$. This is the case for the Wheeler-DeWitt equation derived
from a renormalizable,  two-dimensional dilaton-gravity theory.
One-loop quantum corrections to the CGHS action  have been
calculated by Russo, Susskind and Thorlacius \cite{RST}.
In the conformal gauge $g_{+-}=-e^{2\rho}/2$, $g_{\pm \pm} =0$,
the one-loop  effective action has the form
\begin{equation}
\label{free}
S=\frac{1}{\pi} \int d^2x \left[ -\frac{1}{\kappa} \partial_+ \chi \partial_-
\chi + \frac{1}{\kappa} \partial_+ \Omega \partial_- \Omega
 +\mu^2 e^{2(\chi -\Omega )/\kappa}
+\frac{1}{2} \sum_{j=1}^N \partial_+ f_j \partial_- f_j \right] ,
\end{equation}
where
\begin{equation}
\chi =\kappa \rho -\frac{\kappa}{2} \Phi +e^{-2\Phi}
\end{equation}
represents a Liouville-type field,
\begin{equation}
\Omega = \frac{\kappa}{2}\Phi +e^{-2\Phi}
\end{equation}
is a rescaled version of the dilaton field $\Phi$,
$f_j$ are conformal scalar fields and the constants $\kappa =
(N-24)/12$ and $\mu^2$ are assumed to be
positive-definite\footnote{The reader is
referred to Ref. \cite{RST}  for the details of the derivation.
The numerical value of $\kappa$ is determined by including the
one-loop contributions from the reparametrization ghosts, dilaton and
conformal fields. The theory is one-loop finite if $\kappa =(N-24)/12$.}.

The Wheeler-DeWitt equation corresponding to this
renormalizable model of dilaton-gravity  has been derived
by Mazzitelli and Russo \cite{MR1993}. It has the form
\begin{equation}
\label{WDW}
\left[ \frac{\kappa}{4} \frac{\partial^2}{\partial \chi^2_0}-
\frac{\kappa}{4} \frac{\partial^2}{\partial \Omega_0^2}  -
\frac{1}{2} \sum_{j=1}^N \frac{\partial^2}{\partial f_{j0}^2}
 -4\mu^2 e^{2(\chi_0 -\Omega_0 )/\kappa} -\kappa -2 \right] \Psi =0 ,
\end{equation}
where $\chi_0$, etc., represent the zero modes of the harmonic
expansion of the fields on the cylinder.
In this analysis it is assumed that the coupling between
the zero and higher-order
 modes is negligible and this is equivalent to invoking
the minisuperspace approximation.
This represents an improvement over the approximation
employed to derive Eq. (\ref{WDWab}), however, since this latter
equation follows  from the
classical action (\ref{stringaction}), whereas
Eq. (\ref{WDW})
follows directly from  the one-loop effective action (\ref{free}).

Eq. (\ref{WDW}) is solved by separating
the wavefunction into its gravitational and matter
components with the ansatz $\Psi = \Phi (\chi_0 , \Omega_0 ) \varphi
(f_{j0})$. The plane waves  $\varphi = \exp \left[ i\sum_j Z_j f_{j0} \right]$
form  a basis for the solutions,
where $Z_j$ are arbitrary constants. By identifying  $(x,y )
\equiv (\chi_0 ,\Omega_0 )$, it is readily seen that $\Phi$
statisfies Eq. (\ref{KG}), where the  superpotential is given by
\begin{equation}
\label{2D1}
\kappa m^2= \frac{Z^2}{2} -\kappa -2 -4\mu^2 e^{2(\chi_0 -\Omega_0 )/\kappa}
\end{equation}
and $Z^2 \equiv \sum_j Z^2_j$ represents the total
momentum eigenvalue of the matter sector.
If $Z^2< 2\kappa +4$, $m^2$ is negative-definite and
a function of $(\chi _0 - \Omega_0)$ only. We may therefore
choose $m_+ =1$ in  Eq.
(\ref{uv}) and this  implies that $\Phi$ satisfies an equation of the
form $\partial^2 \Phi /\partial \alpha \partial \beta =-\Phi$,
where
\begin{eqnarray}
\label{beta2d}
\alpha =\lambda (\chi_0 + \Omega_0 ) \nonumber \\
\beta = \frac{1}{\lambda\kappa} \left[ \left( \frac{Z^2}{2} -\kappa -2 \right)
 \left( \chi_0 -\Omega_0 \right)  -2\kappa \mu^2 e^{2(\chi_0 -
\Omega_0 )/\kappa} \right] .
\end{eqnarray}
On the other
hand, the superpotential vanishes along the null line
\begin{equation}
\chi_0-\Omega_0 =\frac{\kappa}{2} \ln \left[
\frac{Z^2-2\kappa -4}{8\mu^2} \right]
\end{equation}
if $Z^2 > 2\kappa +4$ and in this case different
coordinate representations must be employed on either side of this line.

We conclude, therefore,  that the wavefunctions discussed in earlier
sections also apply to this renormalizable model of dilaton-gravity. In
particular,  Lorentzian
solutions to Eq. (\ref{WDW}) may be generated from linear
superpositions of Euclidean solutions and vice-versa when
$\alpha >0$ and $\beta <0$. It follows immediately from Eq. (\ref{beta2d})
that these conditions are satisfied for all $\chi_0 >\Omega_0 >0$ when
$\lambda >0$ and $Z^2<2\kappa +4$.

This concludes our discussion on exact bosonic wavefunctions. In the following
Section  we shall investigate  whether  supersymmetric extentions
to the quantum models discussed above can be performed.

\section{Supersymmetric quantum cosmology}

\setcounter{equation}{0}

\def\theequation{\thesection.\arabic{equation}}

Graham  discovered  that a hidden symmetry exists in
the Bianchi IX Universe by showing how the classical superspace Hamiltonian
may be viewed as the bosonic part of a
supersymmetric Hamiltonian \cite{graham}. It has now been
shown that this hidden symmetry exists
in all Bianchi  class A models \cite{A,susy,review,II}. This implies that a
supersymmetry can be introduced at the quantum level.
This supersymmetric extension of
the quantum theory has significant consequences for quantum
cosmology, as shown by  calculations for  the Bianchi
II Universe \cite{II}. It is thought that these extensions may
provide valuable insight into some of the questions relevant to
a complete theory of quantum gravity. In particular, they may resolve
the problems
encountered when one attempts to  construct a conserved probability from the
wavefunction \cite{review}.
It is therefore of interest to investigate
whether hidden supersymmetries
exist in  two-dimensional dilaton-gravity models.

We begin by briefly reviewing the `hidden symmetry' method of
Graham \cite{graham}. In the minisuperspace approximation
the classical Hamiltonian constraint takes the form
\begin{equation}
\label{bosham}
2H_0= G^{\mu\nu} p_{\mu}p_{\nu} +W(q) =0 ,
\end{equation}
where $G_{\mu\nu}$ is the metric with signature
$(-,+,+,\ldots )$ on the $(D+1)$-dimensional
minisuperspace spanned by the
finite number of degrees of freedom $q^{\mu}$ $(\mu =0,1,\ldots , D)$.
The  momenta conjugate to these variables are $p_{\mu}$ and $W$
represents  the superpotential. This Hamiltonian
is the bosonic component of a supersymmetric
Hamiltonian \cite{witten,rules} if
there exists a function $I(q)$ that
respects the same symmetries as $H_0$
and is itself a solution to the  Euclidean Hamilton-Jacobi equation
\begin{equation}
\label{EucHJ}
W=G^{\mu\nu} \frac{\partial I}{\partial q^{\mu}} \frac{\partial I}{\partial
q^{\nu}}  .
\end{equation}

Fermionic degrees of freedom $\varphi^{\mu} ,\bar{\varphi}^{\nu}$ obeying
 the spinor algebra
\begin{equation}
\label{spinor}
[\varphi^{\mu},\varphi^{\nu} ]_+ = [ \bar{\varphi}^{\mu},
\bar{\varphi}^{\nu} ]_+ =0, \qquad
[ \varphi^{\mu} , \bar{\varphi}^{\nu} ]_{+} =G^{\mu\nu}
\end{equation}
are then introduced. It follows that the supercharges
\begin{equation}
\label{superQ}
Q\equiv \varphi^{\mu} \left( p_{\mu} +i\frac{\partial I}{\partial
q^{\mu}} \right),
\qquad \bar{Q}\equiv \bar{\varphi}^{\mu} \left(
p_{\mu} -i \frac{\partial I}{\partial q^{\mu} } \right)
\end{equation}
satisfy
\begin{equation}
\label{Q}
Q^2=\bar{Q}^2=0
\end{equation}
and, if Eq. (\ref{EucHJ}) is
satisfied, the Hamiltonian (\ref{bosham}) may be written as
\begin{equation}
\label{HQ}
2H_0=[ Q,\bar{Q} ]_+ , \qquad
[H_0,Q]_- =[H_0,\bar{Q} ]_- = 0 .
\end{equation}
These equations represent the algebra
for a single, complex  supersymmetry charge $Q$ and the model therefore
exhibits an $N=2$ supersymmetry \cite{rules}.

This symmetry is preserved at the quantum level
by choosing  the representation
$\bar{\varphi}^{\mu}=\theta^{\mu}$ and $\varphi^{\mu} =G^{\mu\lambda}
\partial /\partial \theta^{\lambda}$ for the
fermionic degrees of freedom, where  $\theta^{\mu}$
are Grassmann variables  \cite{graham,susy}.
The bosonic degrees of freedom have the  usual representation
$p_{\mu}=-i\hbar \partial /\partial q^{\mu}$.
Eqs. (\ref{Q}) and (\ref{HQ}) now represent the
operator realizations of the supersymmetric algebra. The quantized superspace
Hamiltonian is given by
\begin{equation}
\label{superham}
H=H_0 +\frac{\hbar}{2} \frac{\partial^2 I}{\partial q^{\mu}\partial q^{\nu}}
[\bar{\varphi}^{\mu} ,\varphi^{\nu} ]_-
\end{equation}
and has an additional term that vanishes in the classical limit. The existence
of this term  suggests that suitable imaginary or complex  solutions
to Eq. (\ref{EucHJ}) will be difficult to find. It follows
that the supersymmetric
wavefunctions are annihilated by the supercharge
operators
\begin{equation}
\label{squareroot}
Q\Psi =\bar{Q}\Psi =0
\end{equation}
 and it is these constraints that represent
the `square roots' of the Wheeler-DeWitt equation.

\subsection{Induced gravity theory}

To investigate whether the two-dimensional
cosmological models considered
in Section 2 exhibit a hidden supersymmetry of the form discussed above,
we must first  identify the symmetries of
the classical Hamiltonian (\ref{Hequation}).
The kinetic part
is invariant under the   simultaneous interchanges $u\leftrightarrow \pm v$.
However, the full Hamiltonian is not necessarily invariant
under this interchange because of the generality of the
dilaton potential. On the other hand, it is symmetric
under both  simultaneous interchanges if the potential is constant,
i.e. if $V\equiv \lambda^2$.  This form of the potential
in Eq. (\ref{conformalaction})
corresponds to the induced gravity action for $q=\sqrt{8}$ and it
is straightforward to verify that the above  symmetries
are equivalent to an invariance under $\alpha \leftrightarrow \pm \beta$.

Evaluation of Eq. (\ref{betaalpha})
implies that $\beta =q^2\lambda^2 e^{-\psi/q}$ and
comparing  Eq. (\ref{constraint}) with Eq.  (\ref{bosham})
implies that the non-zero components of the minisuperspace
metric are $G_{\alpha\beta}=
G_{\beta\alpha}=a\lambda^2$.
The superpotential is therefore given by $W=-2a\lambda^2$ and it follows
that one solution to Eq. (\ref{EucHJ})
that respects the symmetries of the Hamiltonian is
\begin{equation}
\label{symaction2}
I= -2i(\alpha \beta  )^{1/2} .
\end{equation}
Since we require this `Euclidean' action
to be real, we must assume  that $\lambda^2 <0$.

The functional
form of the supersymmetric wavefunction
may now be determined by solving the constraints (\ref{squareroot}).
Due to the anticommutation
relations obeyed by the Grassmann variables
$\theta^{\mu}$, the general supersymmetric wavefunction
may be expanded as
\begin{equation}
\Psi = A_++B_0\theta^0 +B_1\theta^1 +C_2 \theta^0
\theta^1 ,
\end{equation}
where the bosonic functions $\{ A_+,B_0,B_1,C_2\}$
are functions of $\{ \alpha ,\beta \}$ only.
The annihilation of the wavefunction by the supercharge operators
then translates into a set of coupled, first-order partial
differential equations:
\begin{eqnarray}
\left[ \frac{\partial}{\partial \alpha} +\frac{\partial I}{\partial \alpha}
\right] A_+=0
\nonumber \\
\left[ \frac{\partial}{\partial \beta} +\frac{\partial I}{\partial \beta}
\right] A_+=0
\nonumber \\
\left[ \frac{\partial}{\partial \alpha} +\frac{\partial I}{\partial \alpha}
\right]
B_1 -\left[ \frac{\partial}{\partial \beta} +
\frac{\partial I}{\partial \beta} \right]
B_0=0 \nonumber \\
\left[ \frac{\partial}{\partial \alpha} -\frac{\partial I}{\partial \alpha}
\right]
B_1 +\left[ \frac{\partial}{\partial \beta} -\frac{\partial I}{\partial \beta}
\right] B_0=0 \nonumber \\
\left[ \frac{\partial}{\partial \alpha} -\frac{\partial I}{\partial \alpha}
\right]
C_2 =0 \nonumber \\
\left[ \frac{\partial}{\partial \beta} -\frac{\partial I}{\partial \beta}
\right] C_2 =0 .
\end{eqnarray}

The solution to these equations is given by
\begin{eqnarray}
\label{supersolution}
A_+=e^{-I} \nonumber  \\
C_2=e^{I} \nonumber \\
B_0 =\frac{\partial F}{\partial \alpha} +F \frac{\partial I}{\partial \alpha}
\nonumber \\
B_1 = \frac{\partial F}{\partial \beta} +F\frac{\partial I}{\partial \beta} ,
\end{eqnarray}
where the function $F=F(\alpha ,\beta )$
is itself a solution to the equation
\begin{equation}
\label{F}
\frac{\partial^2 F}{\partial \alpha
\partial \beta} + \left( \frac{\partial^2 I}{\partial
\alpha \partial \beta} -
\frac{\partial I}{\partial \alpha}\frac{\partial I}{\partial \beta }
 \right) F =0 .
\end{equation}

When $I$ is given by Eq. (\ref{symaction2}), Eq. (\ref{F}) simplifies to
\begin{equation}
\label{F1}
\left[ \frac{\partial^2}{\partial w^2} - \frac{\partial^2}{\partial z^2}
-w^2 + z^2 +2 \right] F =0 ,
\end{equation}
where $\{w,z\}$ are defined  by Eq. (\ref{wzdef}).
Hence, $F$ may be interpreted physically as the wavefunction
for a quantum system describing
 two coupled harmonic oscillators that have identical
frequencies but a difference in energy of $2$.
The solution to Eq. (\ref{F1}) has the separable form
\begin{equation}
\label{Fsolution}
F=H_n \left( \sqrt{\gamma\beta} +\sqrt{\alpha} \right)
H_{n+1} \left( \sqrt{\gamma\beta}
-\sqrt{\alpha} \right) e^{-(\alpha +\gamma\beta )} .
\end{equation}

The functions $A_+$ and $C_2$ represent the empty and filled fermion sectors
of the Hilbert space. Both may be interpreted as lowest-order,
WKB approximations to exact  solutions of the bosonic
Wheeler-DeWitt equation (\ref{WDWsr}).  This equation may be
written formally as $\hat{H}_{(0)} \Psi = \hat{H}_{(1)} \Psi $,
where we have split the Wheeler-DeWitt operator into the  two components
\begin{equation}
\hat{H}_{(0)} =-\frac{\partial^2}{\partial s^2} -9\gamma e^{6s}, \qquad
\hat{H}_{(1)} = -\frac{\partial^2}{\partial r^2} .
\end{equation}
Application of $\hat{H}_{(1)}$ implies that $\hat{H}_{(1)} \Psi
=E\Psi$, where $E=p^2$. If the eigenvalue $p$  is real, this equation may be
interpreted as the Schr\"odinger equation, where $E$ represents the
energy associated with $\hat{H}_{(1)}$ \cite{Z}.
Therefore, the state with $E=0$ corresponds to  the state of lowest energy.
When $\beta <0$ and $p=0$,
the general form of the bosonic wavefunction (\ref{solutionsr})
is given by a linear
combination of modified Bessel functions $I_0(x)$ and $K_0(x)$,
where $x= 2\sqrt{\alpha \gamma\beta}$. For large $x$ these functions
have the asymptotic forms $I_0 \propto e^x$ and
$K_0\propto  e^{-x}$,
respectively,  and these limits correspond to the solutions $C_2$ and $A_+$.
The supersymmetric vacua are
therefore closely related to their  semi-classical
limits and correspond to the pure bosonic states of lowest energy.
These features appear
 to be  generic properties  of supersymmetric ground state
wavefunctions \cite{rules}.

\subsection{A string inspired model}

String theory exhibits a symmetry known as target space duality
\cite{duality}. (For a recent review see, e.g., Ref.
\cite{dualityreview}). In two-dimensional space-times, a string cannot
tell if it
is propagating on a circle of radius $a$ or of radius $a^{-1}$. In effect,
this allows one to transform between theories of radii $a$ and $a^{-1}$ after
a suitable translation on the dilaton field \cite{c}.
It is convenient to consider this symmetry within the context
of   the action (\ref{stringaction})
in the absence of loop corrections, i.e. $D=c>0$.
The Hamiltonian derived from this action
 for the cosmological space-time  (\ref{line}) is given by
\begin{equation}
\label{stringham}
H=e^{-2\Phi} \left[ \frac{4}{N^2} \left(
\dot{a}\dot{\Phi} -a\dot{\Phi}^2 \right)
 -c a \right]
\end{equation}
and it is straightforward to verify that  Eq.
(\ref{stringham}) is invariant under the duality transformation
\begin{equation}
\label{duality}
a=\frac{1}{A}, \qquad \Phi = \phi +\ln a .
\end{equation}

If we introduce the coordinate pair
\begin{equation}
\label{XYdef}
X\equiv ae^{-\Phi}, \qquad Y=e^{-\Phi} ,
\end{equation}
the Hamiltonian takes  the  form
\begin{equation}
\label{stringham2}
H=-\frac{4}{N^2} \dot{X}\dot{Y} -cXY
\end{equation}
and  invariance under the duality transformation (\ref{duality})  is
therefore equivalent
to an invariance under the simultaneous interchange $X \leftrightarrow Y$.

The momenta conjugate to $X$
and $Y$ are $p_X=4\dot{Y}/N$ and $p_Y=4\dot{X}/N$, respectively.
It is convenient to perform a rescaling of these degrees of  freedom
by defining $\alpha \equiv X^2$ and $\beta \equiv Y^2$. The classical
Hamiltonian (\ref{stringham2})
is then given by Eq.  (\ref{bosham}), where the non-vanishing
components of the minisuperspace metric are $G_{\alpha\beta}=
G_{\beta\alpha}=(\alpha\beta)^{1/2}$ and the
superpotential $W=2c(\alpha\beta)^{1/2}$.
There exists a hidden supersymmetry if $I$ satisfies
\begin{equation}
\label{EucHJstring}
\frac{\partial I}{\partial \alpha}\frac{\partial I}{\partial \beta} =c
\end{equation}
and also respects the duality symmetry $\alpha \leftrightarrow \beta$.
One solution satisfying the necessary conditions
is $I=2\sqrt{c\alpha\beta}$ and since $\{ \alpha ,\beta \}$ are
positive-definite, $c>0$ is necessary for the solution to be real.

We conclude, therefore, that
the general supersymmetric wavefunction may also be found
in closed form for this theory.
The bosonic functions in the Grassmann basis
expansion of the wavefunction are again given by Eq. (\ref{supersolution}),
but $F$ has the  slightly different form
\begin{equation}
F=H_{n} (\eta ) H_{n+1} (\xi ) e^{-(\xi^2 + \eta^2 )/2} ,
\end{equation}
where $\xi \equiv c^{1/4}(\alpha^{1/2} +\beta^{1/2} )$ and
$\eta \equiv c^{1/4}(\alpha^{1/2} -\beta^{1/2} )$.

\section{Conclusions and discussion}

\setcounter{equation}{0}

\def\theequation{\thesection.\arabic{equation}}

In this paper we have investigated the quantum
cosmology of a generalized class of two-dimensional
dilaton-gravity models. If the dilaton potential contains
no roots, i.e. if $V(\psi) \ne 0$ for all physically
interesting $\psi$, the classical
dynamics of these Universes is equivalent to that  of a
non-interacting, point particle propagating over a portion of
two-dimensional Minkowski space. A large subset of this class
of models is dynamically equivalent to
the isotropic, constrained oscillator-ghost-oscillator system.
This suggests that the relationship between
quantum configurations and classical space-times, as
discussed in Ref. \cite{OT},  could be generalized to these models.

Furthermore,
these correspondences imply that the Wheeler-DeWitt equation can be expressed
as the unit-mass Klein-Gordon equation if a suitable choice of factor
ordering is made. This allows a number of exact and approximate
solutions to be found. Quantum states corresponding to
Lorentzian geometries may be generated from an infinite sum of Euclidean
solutions and vice-versa. The
Hamilton-Jacobi equation can be  solved by employing a Legendre  transformation
and all developable solutions to this equation were found in parametric form.

We proceeded to identify a wider class of integrable two-dimensional
minisuperspaces that can be solved exactly by
mapping the Wheeler-DeWitt equation  onto the unit-mass Klein-Gordon
equation. This mapping is possible if the superpotential
of the wavefunction is a separable function of the null
coordinates over minisuperspace. We applied this result
to the Wheeler-DeWitt equation derived from a renormalizable,
two-dimensional dilaton-gravity model \cite{MR1993}.

One of the main problems with the quantum cosmology program is
the construction of a non-negative norm from solutions to the
Wheeler-DeWitt equation. This equation is a hyperbolic, second-order partial
differential equation,  so the conserved current associated with
it is not necessarily semi-positive definite. Consequently, it
is not clear that such  a current will provide a suitable measure
of probability. A similar problem
is encounted when the Klein-Gordon  equation for a scalar
field is solved. In this case, however, the ambiguity is resolved by
taking the `square root' and
in view of the close correspondence between the Wheeler-DeWitt and
Klein-Gordon equations, it has been suggested that
a similar technique might solve the corresponding problem
in quantum cosmology \cite{review,bb}.

This suggests that one should search for supersymmetric extensions
to quantum cosmology. It was shown in Section 6
that the classical Hamiltonians
derived from the induced gravity theory  and a string-inspired model may be
viewed as  the bosonic components of a supersymmetric Hamiltonian. In
the latter case, the origin of this symmetry may  be traced to
the invariance of  string theory under duality transformations.
The hidden symmetry method was employed to derive
the corresponding quantum constraints for the two models.
This method differs from other approaches to supersymmetric
quantum cosmology because it does not start from  a field theory of
supergravity \cite{II}. The quantum
constraints can be solved exactly and closed-form  expressions
for  the general supersymmetric wavefunction were found. It would be
interesting to investigate whether this method can be applied
to more general models.

An alternative approach to quantum cosmology is
the third quantization procedure \cite{third,formal}.
The aim of this approach is to
construct a consistent probabilistic measure in quantum gravity by
promoting the wavefunction of the Universe  to
a quantum field operator that acts on a Hilbert space
of states. The  `vacuum' state in this space is identified as the state
where the Universe does not exist. Topology  changing processes can then
be described by including self-interactions of the Universe field.
Moreover, in the  minisuperspace  approximation
a suitable combination of the dynamical
degrees of freedom may be associated with a time variable
in the Wheeler-DeWitt equation. It then follows that the superpotential
of the wavefunction may be viewed as a  `time-dependent' function.
In ordinary quantum field theory it
is well known that particles are
created from the vacuum by a time-varying external  potential and
this suggests that Universes  could be created via a similar process.
In practice   the Universe field is expanded  into
positive frequency in- and out-mode functions and their hermitian conjugates.
The in- and out-modes
are related to one another by the Bogoliubov coefficients
and these determine the number of Universes in a given mode \cite{BD}.
The creation of Universes in this picture
arises because  the two Hilbert  spaces generated by the in- and out-mode
functions
are inequivalent and this  results in non-zero Bogoliubov coefficients.

Recently Vilenkin \cite{vil} has argued against this picture of
Universe creation. His main objection is that the time variable constructed
in minisuperspace models is generally not a monotonically
increasing  function since
Universes can contract as well as expand. He then interprets the creation
of a pair of Universes in terms of a contracting Universe that
 begins reexpanding at a finite radius.
On the other hand, he does suggest that third quantization might be
appropriate for describing topology changing processes  in two-dimensional
Universes.

Since there is currently no generally accepted interpretation
of third quantization, it is of interest to investigate its
consequences further. The procedure can be applied to the class of
two-dimensional models (\ref{conformalaction}) for which $\beta$, as defined
in Eq. (\ref{betaalpha}), is positive-definite for all $\psi$.
The variables (\ref{srdef}) take
all values in the range $(s,r)\in (-\infty ,+\infty )$ and we
may therefore view $s$ as the time variable in the Wheeler-DeWitt equation
(\ref{WDWsr}). The scale factor of the Universe
vanishes  as $s\rightarrow -\infty$ and infinite spatial volume
corresponds to the limit $s\rightarrow \infty$.
Formally, this model is identical to the
one considered previously by Hosoya and Morikawa \cite{formal}, so
their results will apply here. The
appropriately normalized positive-frequency in- and
out-mode functions are given by
\begin{eqnarray}
u_p^{\rm in} (s,r) =\left( \frac{\pi}{6} \right)^{1/2}
\left( {\rm sinh} \frac{\pi|p|}{3} \right)^{-1/2}
e^{ipr} J_{\nu} \left( e^{3s} \right) \nonumber \\
u_p^{\rm out} (s,r) =\frac{1}{2} \left( \frac{\pi}{3}
\right)^{1/2} e^{-\pi|p|/6} e^{ipr} H_{\nu}^{(2)} \left( e^{3s}
\right) ,
\end{eqnarray}
respectively,
where $\nu =-i|p|/3$. As $s\rightarrow \infty$, $u_p^{\rm out}
\propto  e^{iS} $ and
this is the WKB solution given by Eq. (\ref{actionsolution}) with
$f=0$.
It follows that the Bogoliubov coefficients are given by
$c_1(p,q) =b_1\delta_{pq}$ and $c_2(p,q)=b_2\delta_{pq}$, where
\begin{equation}
b_1 =\left( 1-e^{-2\pi |p|/3} \right)^{-1/2},
\qquad |b_2| =\left( e^{2\pi|p|/3} -1 \right)^{-1/2}
\end{equation}
and  the average number of Universes  with `energy'
$p$ therefore has a Planckian distribution
\begin{equation}
N_p = |c_2(p,p)|^2 =\left( e^{2\pi|p|/3}-1 \right)^{-1} .
\end{equation}

Hosoya and Morikawa \cite{formal}
extended this free field theory by including a $\Psi^3$ interaction that
describes the spliting of a `mother' Universe into
two `baby' Universes of identical topology. By treating  the mother
Universe in a classical fashion,
they showed that the quantized baby Universes also
have a Planckian distribution.
The formal equivalence of their model with those studied in this work
suggests that similar conclusions should apply for a wide class of
two-dimensional cosmologies. It would be of interest to investigate
these possibilities further.

\vspace{.7in}

The author is supported by the Particle Physics and Astronomy Research
Council (PPARC), UK.

%%%%%%%%%%%%%%%%%%%%%%%%%%%%%%%%%%%%%%%%%%%%%%%%%%%%%%%%%%%%%%
%%%%%%%%%%%%%%%%%%%%%%%
\frenchspacing
\def\prl#1#2#3{{ Phys. Rev. Lett.} {\bf #1}, #2 (#3)}
\def\prd#1#2#3{{ Phys. Rev. D} {\bf #1}, #2 (#3)}
\def\plb#1#2#3{{ Phys. Lett. B} {\bf #1}, #2 (#3)}
\def\npb#1#2#3{{ Nucl. Phys. B} {\bf #1}, #2 (#3)}
\def\apj#1#2#3{{ Ap. J.} {\bf #1}, #2 (#3)}
\def\apjl#1#2#3{{ Ap. J. Lett.} {\bf #1}, #2 (#3)}
\def\cqg#1#2#3{{Class. Quantum Grav.} {\bf #1}, #2 (#3)}
\def\grg#1#2#3{{Gen. Rel. Grav.} {\bf #1}, #2 (#3)}
\def\mnras#1#2#3{{Mon. Not. R. astron. Soc.} {\bf #1}, #2 (#3)}
%%%%%%%%%%%%%%%%%%%%%%%%%%%%%%

%%%%%%%%%%%%%%%%%%%%%%%%%%%%%%%%%%%%%%%%%%%%%%%%%%%%%%%%%%%%%%%%

\end{document}